\def \be {\begin{equation}}
\def \ee {\end{equation}}

\def \ts {\textstyle}
\def \bea {\begin{eqnarray}}
\def \eea {\end{eqnarray}}
\def \d {\mbox{d}}
\documentstyle[12pt]{ioplppt}

\begin{document}

\title{Exact isotropic cosmologies with local fractal 
number counts} 

\author{Neil P. Humphreys, David  R. Matravers
and Roy Maartens }

\address{School of Computer Science and Mathematics, 
University of Portsmouth, Portsmouth PO1 2EG, Britain}


\begin{abstract}
We construct an exact relativistic cosmology in which an 
inhomogeneous but isotropic local 
region has fractal number counts and matches to a homogeneous 
background at a 
scale of the order of $10^2$ Mpc. We show that Einstein's equations 
and the matching conditions imply either a nonlinear Hubble law
or a very low large-scale density. 

\end{abstract} 

\pacs{9880H, 9880E, 0420J, 0440N}

\section{Introduction}

There is doubt and argument about whether the data on the galactic 
number count $N$ can support what is called a fractal structure, i.e.,
 \be
N \propto y^{\nu}\,,   \label{fractal}
 \ee
where $0 < \nu < 3 $ is the fractal index and $y$ 
is some distance measure (see \cite{pee,pie,syl} and references 
therein).  However, there is evidence for such a pattern in the 
number counts out to a distance of the order of
$10^2h^{-1}$ Mpc \cite{pee,pie,mon}. 

Even this evidence is beset with problems because it is difficult
to agree adequate statistics which are model independent
\cite{pie}.  For instance, Peebles (\cite{pee}, page 212) 
specifically 
excludes inhomogeneous spherically symmetric models in which we are 
at 
the centre. This may be reasonable but it serves to illustrate how 
aware one needs to be of the underlying model.  It would help if good 
predictive dynamical models for the local distribution were 
available. 
The data could then be tested against these.  Attempts in this 
direction are in their infancy but there are some that could yield a 
fractal number count \cite{veg}. 
At the same time, a number of workers are using recent 
large-scale structure observations and indirect evidence to
argue that fractal-like clustering on small and possibly
intermediate scales does not continue indefinitely, and that
there is a transition to a near-homogeneous distribution
on large scales \cite{anti}. This is supported by the implications
for large-scale structure of the near-isotropy of the cosmic 
microwave background radiation, since it has been shown that
if all fundamental observers measure small anisotropy in the
temperature, then the universe on large scales after last scattering 
is necessarily close to homogeneity \cite{cmb}.

Our study will consider the number counts as a function of the 
observer area distance (angular diameter distance)
$D$ or redshift $z$, down the past light cone 
of the observer.  As a consequence, even in the spatially homogeneous 
Friedman-Lema\^{i}tre-Robertson-Walker (FLRW) case the number counts 
will be non-homogeneous.  However the underlying spatial homogeneity 
of the FLRW geometry is reflected in the fact that the number 
count/redshift formula is precisely determined by the model 
\cite{ell}.  This formula specifically rules out a simple fractal 
distribution of the form (\ref{fractal}).
Note that our approach precludes direct comparison with 
work in which the fractal distribution is assumed to hold on 
spacelike surfaces (see for example \cite{ras}). 

To analyse the problem theoretically we assume (a) that locally we
are in a spherically symmetric region which can be modelled by a 
Lema\^{i}tre-Tolman-Bondi (LTB) geometry with a fractal number count 
of sources down our past light cone, and (b) that at some value of 
the 
observer area distance (of order $10^2h^{-1}$ Mpc) 
the universe becomes 
homogeneous and can be modelled by an appropriate FLRW spacetime.  We 
assume that the LTB region matches smoothly to the FLRW region, with 
no surface density layer or shell crossing \cite{jmp,hellaby}. These 
assumptions are not unreasonable on the basis of current 
observations. 
Of course, we are only considering a very restricted form of
fractal distribution, i.e. a single fractal in the radial direction
which maintains the overall spherical symmetry.

A similar model was investigated by
Ribeiro \cite{rib}, who was the first to construct a relativistic
fractal-count model. Ribeiro developed sophisticated numerical
computations to analyze his models. He showed that parabolic
and elliptic models gave a nonlinear Hubble law, while hyperbolic
models were able to support a linear Hubble law. Our results
are in agreement with these conclusions, but we believe that
we have identified more clearly and simply the underlying
reasons for these features. We have also pursued the implications
of parabolic models that match to open FLRW models -- i.e. that
the large-scale density is extremely low.
Ribeiro's work concentrated on numerical integration.
Our approach is primarily analytical, and relies crucially on
a systematic use of the central regularity conditions and
the transition matching conditions. Thus our work may be
seen as complementing Ribeiro's by bringing analytical insights to
some of his results and extending aspects of his work.

In principle, LTB models can accomodate a great variety of number
count relations \cite{ell,maa,mhe}. However, the assumed number count
relation may not be compatible with matching to the large-scale
homogeneous universe, with other observations, or with regularity
conditions on the observer's wordline.

The main result here is that if the fractal number counts are taken 
at 
face value and the universe can be modelled locally by an LTB 
geometry 
which matches to an FLRW background at some scale,  then the universe 
either cannot have a linear Hubble law at low redshift (parabolic
case), or it
has a very low large-scale density (non-parabolic case).  
Perturbation of the model is unlikely to 
change this conclusion and increasing the scale of the fractal-count 
region 
(and there is some evidence for this) will exacerbate the low density
problem.  The low density problem 
identified by this model can only be avoided if the dark matter is 
strongly biased. 

It would be of interest to construct a linear perturbative model 
to confirm the above.  The objective here, however, is to 
use an exactly isotropic non-perturbative model and to study the 
consequences flowing from the field equations and matching 
conditions.

\section{The cosmological model}

In this section equations are presented for the general LTB metric, 
which, together with a Kantowski-Sachs-type solution \cite{jmp},
encompasses all regular, spherically symmetric dust models.
These equations are used in section 3 to analyse the fractal 
number count subclass of LTB models. 

The metric must satisfy regularity conditions, including the central 
conditions as $D\rightarrow 0$. (See \cite{ell} for the central 
conditions in general;
a comprehensive analysis 
for LTB is given in \cite{jmp} for central observers, and in
\cite{hmm} for off-centre observers.) For this problem we use 
coordinates based on the observer's past light cones as described in 
\cite{ell} and usually called observational coordinates. 
Details of the constuction for the LTB model are given in \cite{maa}. 
(See also \cite{mm}, where semi-isotropic observational
generalizations of LTB models are investigated.)

Explicitly the coordinates are  $\{w,y,\theta,\phi\}$,  where
$\{w=$constant$\}$ are the observer's past light cones, $y$ is a 
distance 
from the observer along a past light ray and $(\theta,\phi)$ 
label the direction of observation.   In these coordinates the metric 
of the LTB model is 
 \[
 \d s^{2} = - A(w,y)^{2}\d w^{2}+2A(w,y)B(w,y) \d w \d y+C(w,y)^{2} 
\left(\d\theta^{2}  + \sin^{2} \theta \d \phi^{2} \right)\,.
  \]

The Einstein field equations cannot be integrated explicitly in these 
coordinates \cite{maa}, but they can be in the more familiar (1+3) 
formalism \cite{tol}.  However the (1+3) coordinates are not directly 
observable.  An advantage of the observational coordinates is that 
the 
null geodesic equations are already integrated, whereas they are not 
integrable in the (1+3) formalism.  This explicit relation between 
the 
null geodesics and the coordinates facilitates the interpretations.  
We will use both formalisms below. The metric in (1+3) comoving
coordinates has the form 
  \[ 
  \d s^2=-\d t^2+
  \left[{\partial R(r,t)\over\partial r}\right]^2
{\d r^2 \over 1-kf(r)^2}
+R(r,t)^2\left(\d\theta^2+\sin^2\theta \d\phi^2\right) \,,
 \] 
where $f$ is arbitrary and relates to the `total energy' of the LTB 
model. The cases $k = 0, +1,-1$ correspond respectively to parabolic, 
elliptic or hyperbolic geometry of the $\{t=$ constant $\}$ 
hypersurfaces. If $\{w = w_{0}\}$ is the past light cone of 
observation, then 
 \be 
D(y(z))=C(w_0,y(z))=R(r(z),T(r(z)))\,,
 \ee
 where $t=T(r)$ is the equation of a past light ray.

In the observational coordinates the number count (total number of 
sources out to distance $y$) is given by \cite{ell,maa}
 \[
 N(y) = 4 \pi \int^{y}_{0} n(w_{0}, x) B(w_{0}, x) 
D(x)^{2} \d x \,,
 \]
where $n(w_{0}, y)$ is the number density of sources.   The 
dust density is 
 \[
 \rho = mn \,,
 \]
where $m$ is the average galactic mass, which we assume to be 
constant.  For simplicity, 
we omit evolution and selection effects. It would be 
possible to take some account of these effects when choosing 
$n$ 
and $m$,  but we have not done so.  

The total number count in the inhomogeneous region is given in terms 
of the Bondi mass function $M(r)$ (which arises from integrating the 
field equations) by \cite{maa}
 \be
 N(y) = \frac{1}{m} \int^{y}_{0} M'(x)\left[1-kf(x)^2\right]^{-1/2} 
 \d x\,,
 \label{count}
 \ee
where $M(y)$ is short-hand for $M(r(y))$.

It is well known that the LTB metric is fully determined by three 
arbitrary functions: one can be removed by a coordinate choice but 
the 
other two have physical significance.  In \cite{maa} quadratures 
are given which determine the LTB functions directly from the 
observational data $D(z)$ and $N(z)$. This determination forms the 
corner-stone of the current
analysis, since the LTB functions are needed to 
determine the dynamics and (crucially) the matching to the FLRW 
model. 
If the LTB functions are identified as the Bondi mass $M$, the energy 
function $f$ and the big-bang time $\beta$ (see \cite{bon}), then we
can use the results of \cite{maa} to give these functions
in terms of the observables as follows: 
 \be
\sqrt{1-kf(z)^2} = 
\frac{(1+z)}{2D(z)}\,\int_0^z{1\over Q}\left\{D'(x)+ 
 \left[\frac{D(x)Q(x)^2} {(1+x)^2}\right]'\right\}\d x\,,
 \label{neil1}
 \ee 
where 
\be 
Q(z) \equiv 1- m \int_0^z\frac{(1+x)N'(x)}{D(x)} 
\,\d x\,. \label{neilly} 
 \ee 
 By a rearrangement of (\ref{count})
\[
 M(z) = m\int_0^z \left[1-kf(x)^2\right]^{1/2} N'(x)\, \d x \,, 
\]
and 
\bea
  \beta(z) & = &  t_0-\int_0^z D'(x) \left\{
  \left[1 - kf(x)^{2}\right]^{1/2} -
\left[\frac{2M(x)}{D(x)} - kf(x)^{2}\right]^{1/2}\right\}^{-1}\d x 
 \nonumber\\ 
 & &{} + \left\{
 \begin{array}{ll} 
 -{\textstyle{1\over3}}\left[2D(z)^3 M(z)^{-1}\right]^{1/2} & ({\rm 
if}~~k=0)\\ 
{}&{}\\
M(z)\left[\sin\Gamma(z)-\Gamma(z)\right]|f(z)^{-3}| & ({\rm 
if}~~k=+1)\\ 
{}&{}\\
M(z)\left[\Gamma(z)-\sinh\Gamma(z)\right]|f(z)^{-3}| &({\rm if}~~k=-
1)\,,
\end{array} 
\right. 
\label{neil3}\eea 
where
 \be
\Gamma(z) \equiv \left\{ 
\begin{array}{ll} 
2\,{\rm arcsin} \left[{\textstyle{1\over2}}D(z) 
f(z)^{2}M(z)^{-1}\right]^{1/2} & ({\rm if}~~k=+1) \\ 
{}&{}\\
 2\,{\rm arcsinh} \left[{\textstyle{1\over2}}D(z)f(z)^2 M(z)^{-1} 
 \right]^{1/2}& ({\rm if}~~k=-1) \,.
\end{array} \right. \label{neilly2}
 \ee
 An arbitrary constant of integration appearing in (\ref{neil3}) has 
been identified as $t_0$, the time of observation
($t_0$ is arbitrary since only the time elapsed after 
the big bang, $t-\beta$, has physical significance). In these 
integrations the freedom in $y$ on the light cone of observation has 
been used up by setting $A(w_0,y)=B(w_0,y)$; this choice simplifies 
the calculation \cite{maa}. 

Now it is evident that in the $k=0$ case only one of the functions 
$N(z)$ and $D(z)$ is arbitrary;  for example $N(z)$ may be obtained 
from (\ref{neil1}) by setting $k=0$, once $D(z)$ is known. This 
covariant constraint on the observational data may be 
found explicitly as follows.
By (\ref{neilly})
\be
1+z=-{D\over m} {\d Q\over \d D}\left({\d N\over\d D}\right)^{-1}\,,
\label{new1}\ee
and then (\ref{neil1}) with $k=0$ gives
\[
Q{\d\over\d D}\left[-2m{\d N\over\d D}\left({\d Q\over\d D}
\right)^{-1}\right]={\d\over\d D}\left\{D\left[1+{Q^2\over
(1+z)^2}\right]\right\}\,.
\]
Integrating by parts, and using the central conditions, we find that
\[
-2mN+2m{\d N\over\d D}\left({\d Q\over\d D}\right)^{-1}+D+
{m^2Q^2\over D}\left({\d N\over\d D}\right)^2\left({\d Q\over\d D}
\right)^{-2}=0\,.
\]
Solving this as a quadratic (and again imposing the central
conditions to eliminate a spurious root), gives
\[
{1\over Q}{\d Q\over\d D}=-{m\over D}{\d N\over\d D}
\left(1-\sqrt{{2mN\over D}}\,\right)^{-1}\,.
\]
Together with (\ref{new1}), this gives the solution
\be
1+z =
\left(1-\sqrt{{2mN\over D}}\,\right)^{-1}
\exp\left[-\int {m\over D}{\d N\over\d D}
\left(1-\sqrt{{2mN\over D}}\,\right)^{-1}
\d D\right] \,,
\label{new2}\ee
which appears to be a new result, and which is central to
showing that the Hubble law is nonlinear (see below).
Thus when $k=0$, if $N$ is known in terms of $D$, then (\ref{new2})
gives $z$ in terms of $D$.

\section{Fractal number counts}

The model we construct is isotropic about the observer 
and inhomogeneous out to a distance   
$D = D_h$ of the order of $10^2$ Mpc, corresponding to 
a redshift $z$ below $10^{-1}$.  
The LTB spacetime 
metric in this local region
satisfies the Einstein equations with a dust source which 
yields a power-law number count with fractal index, of the 
form (\ref{fractal}).
To ensure 
consistency with observational data, we 
also require a linear distance/redshift relation out to 
the homogeneity scale $D_h$,
i.e. a linear Hubble law for small redshift. The Hubble constant is
$H_0=100h$ km s$^{-1}$ Mpc$^{-1}$, with $0<h<1$.
For the large-scale universe $(D>D_h)$, we 
require a homogeneous FLRW geometry. The Darmois matching 
conditions at $D_h$ uniquely determine the parameters of this FLRW 
solution. 

In all cases the central conditions \cite{maa} demand that 
for small $D$ 
 \be
N(D) = \left(\frac{4 \pi \rho_{0} }{3m} \right) D^{3} + 
O(D^{4})\,,  \label{cube}
 \ee
so that $N \sim D^{3}$ as $D\rightarrow 0$.  
This means that the number count 
cannot be fractal (i.e. with $\nu\neq 3$) for $D$ near zero.  There
is some minimum distance $D_f$ for fractal counts, 
below which $N\sim D^3$. (Note that
$D_f$ should be above the averaging scale which is implicit in
a continuum dust model of galactic matter.)
A simple model of fractal number counts that incorporates the
limiting behaviour (\ref{cube}) is the continuous power-law
ansatz 
 \be
 N(D)  = \left\{ 
 \begin{array}{ll} 
 \left({4\pi\rho_0\over 3m}\right) D^{3} & 
 {\rm for} ~~ D \leq D_f  \\
 {}&{}\\
 \left({4\pi\rho_0\over 3m}\right) D_f^{3}\left({D\over D_f}
 \right)^\nu & 
  {\rm for } ~~ D_f \leq D < D_h \,.
   \end{array} 
\right.
\label{match} 
 \ee
(Matching conditions require $N$ to be at least continuous at 
$D = D_f$, otherwise there would be a surface layer.)  
This model represents an `instantaneous' transition from
non-fractal to fractal counts at the minimum fractal distance
$D_f$, and $N$ is not differentiable at $D_f$.
An alternative model, in which $N$ is a smooth function of $D$,
is
  \be 
 N =\left({4\pi\rho_0\over 3m}\right)D^3\left(1 + \frac{D}{D_f} 
\right)^{\nu - 3} \,,
\label{an2}
\ee
which satisfies $N\sim D^{3}$ for $D \ll D_f$ 
and $N \sim D^{\nu}$ for $D \gg D_f$. 

Since 
$D_f$ has to be small, the region where $D < D_f$ can be treated 
as homogeneous and so $z_{f} \equiv z(D_f) \approx H_{0}D_f$.  
At the homogeneity scale $D_h$, the number counts must match 
continuously to the FLRW number count relation $\bar{N}(D)$:
\be
N(D_h)=\bar{N}(D_h) \,.
\label{new3}\ee
The functions $\bar{N}(D)$ are known. For example, the $k=0$
FLRW spacetime has \cite{maa}
\be
H_0D=2
\left({mH_0\over 32\pi}\,\bar{N}\right)^{1/3}\left[1-
\left({mH_0\over 32\pi}\,\bar{N}\right)^{1/3}\right]^2 \,.
\label{new4}\ee

\subsection{Parabolic fractal-count models}

Consider now the possibility of modelling the fractal-count region 
$D_f < D < D_h$ by a parabolic LTB solution ($k=0$). Assuming
that the core region $0\leq D<D_f$ is also parabolic, 
Einstein's equation determines the Hubble constant
in terms of the central density (as in a FLRW model) \cite{maa}:
\[
H_0=\sqrt{{\ts{8\over3}}\pi\rho_0}\,.
\]
We assume that the fractal number counts are modelled by the power-law
ansatz (\ref{match}).
Then the matching of $N$ at the homogeneity scale $D_h$, given
by (\ref{new3}) and (\ref{new4}), implies
\be
\left[H_0D_f\left({D_h\over D_f}\right)^{\nu/3}-2\right]^6-
8\left({D_h\over D_f}\right)^{3-\nu}=0 \,.
\label{new5}\ee
Thus the four parameters $H_0$, $D_f$, $D_h$ and $\nu$ are subject
to the constraint (\ref{new5}) by virtue of number count continuity.
If observations are used to determine $H_0$, $D_h$ and $\nu$, then
this constraint fixes the minimum fractal scale $D_f$.

The problem with the parabolic fractal-count models arises from
the nonlinear behaviour of the redshift/ area distance relation
for small $D$. From equations (\ref{new1}) and (\ref{new2}),
we find that (\ref{match}) implies
\be
{\d z\over\d D}\approx{\ts{1\over2}}H_0\left[(\nu-1)\left({D_f\over D}
\right)^{(3-\nu)/2}-\nu H_0D_f\left({D_f\over D}\right)^{2-\nu}
\right] \,,
\label{new6}\ee
for small $D$. It follows that
\[
{\d z\over\d D}\sim \left\{ \begin{array}{ll}
-D^{(\nu-3)/2} & \mbox{ for } 0< \nu < 1 \\
{}&{} \\
-D^{\nu-2} & \mbox{ for } 1\leq \nu<3 \,.
\end{array}\right.
\]
Thus after the initial linear behaviour up to $D_f$ (by construction),
the redshift/distance graph begins immediately
to curve downwards.
This nonlinearity contradicts the
well-established linear Hubble law on scales up to the order of $10^2$
Mpc, and means that parabolic fractal-count models are ruled out.
Although we have used the power-law fractal count ansatz (\ref{match})
to deduce this nonlinearity, it is clear that the feature will
persist for any model that incorporates (\ref{fractal}), since the
argument depends only on the behaviour for small $D$.

Finally, we note that exact expressions for $z(D)$ may be
obtained
for any rational fractal index $\nu$, since in this
case the quadrature in (\ref{new2})
may be performed exactly for the power-law relation (\ref{match}). 
For example, with $\nu = \frac{3}{2}$ we find
 \be
1 + z = (1 + z_{f}) \left[\frac{1 - H_{0}D_f(D/D_f)^{1/4}}{1 - 
H_{0}D_f} \right]^{2} \exp\left[3H_{0}D_f\left\{\left({D\over
D_f}\right)^{1/4} - 1\right\}\right] \,, \label{89}
 \ee
for $ D_f < D < D_h $. 
Nonlinearity for small $D$ is readily confirmed by plotting
the graph of (\ref{89}).

\subsection{Non-parabolic fractal-count models}

The interpretation of observations cannot be realised if the 
fractal-count
region is parabolic, since then the redshift/distance relation is 
nonlinear. For $k=\pm 1$ there is no such problem, since (as 
discussed 
in section 2) the observable functions $N(z)$ and $D(z)$ are 
independent. This allows us to circumvent the nonlinearity
problem -- but a new problem arises, i.e. the problem of very low
density in the large-scale FLRW region arising from fractal
number counts and the matching conditions. Intuitively, fractal
number counts imply an under-density (since $\nu<3$), so that 
$k=-1$. Then matching conditions rule out $k=+1$ on large
scales, and imply that the large-scale $k=-1$ FLRW model
must have even lower density than the local LTB region.

From (\ref{count}) the relationship between $M$ and $N$ 
is more complicated in LTB models with $k\neq0$. If $k=+1$, one 
obtains 
$M/m<N$, and conversely for $k=-1$.  Note that regularity in the 
current context dictates that $k$ must not increase with distance 
from 
the observer \cite{jmp}. Therefore we cannot match a $k = -1 $ LTB 
model to a $k = +1 $ LTB or FLRW exterior.  Furthermore, the matching 
conditions require continuity of $M$ and $kf^2$. 

Numerical integrations are simpler with differentiable 
functions. For this purpose we replace (\ref{match}) with the
alternative smooth-transition ansatz (\ref{an2}).
We now focus on the dynamics and 
the implications of such a number count profile (with the other 
observations)
for the large-scale
density.  
Using the formulas (\ref{neil1})--(\ref{neilly2})
we have integrated the field equations with number count formula 
(\ref{an2}) out to $z = 0.07$,  where we assume the metric matches to 
the FLRW background.  
We take $D_f = 10$ Mpc.
We label the 
central (local) density parameter by $\Omega_{c}$ and the large-scale 
(background) density parameter by $\Omega_{0}$. We take $D(z) = 
H_{0}^{-1}z$, which is the well established Hubble law for these 
redshifts. In 
the following tables the remaining parameters are varied in turn and 
the consequences for the background density are shown. \\

\[ \]

\begin{center}
Solutions with $h=0.65$, $\Omega_c = 0.2$
\[ \]
\begin{tabular}{|l|l|}\hline
fractal & large-scale \\
index $\nu$ & density $\Omega_{0}$\\
{}&{}\\\hline\hline 
1.0 & 0.0002 \\\hline
1.5 & 0.001  \\\hline
2.0 & 0.008  \\\hline
2.5 & 0.04   \\\hline
3.0 & 0.2   \\\hline
\end{tabular}\\

\end{center}
\[ \]
\begin{center}
Solutions with $h=0.65$, $\nu=1.5$
\[ \]
\begin{tabular}{|l|l|}\hline
local & large-scale \\
density $\Omega_c$ & density $\Omega_{0}$\\
{}&{}\\\hline\hline 
0.20 & 0.001 \\\hline
0.35 & 0.0025  \\\hline
0.50 & 0.004  \\\hline
0.75 & 0.006   \\\hline
1.00 & 0.008   \\\hline
\end{tabular}\\

\end{center}
\[ \]
\begin{center}
Solutions with $\nu=1.5$, $\Omega_c=0.2$
\[ \]
\begin{tabular}{|l|l|}\hline
Hubble & large-scale \\
rate $h$ & density $\Omega_{0}$\\
{}&{}\\\hline\hline 
0.20 & 0.00025 \\\hline
0.40 & 0.0007  \\\hline
0.65 & 0.001  \\\hline
0.80 & 0.002   \\\hline
1.00 & 0.0025   \\\hline
\end{tabular}
\end{center}
\[ \]
\newpage

In all of these cases it was found that $k$ had to be equal $-1$, 
corresponding to hyperbolic space geometry as one would expect with 
low densities.  The background density is that of the uniquely 
defined 
FLRW model that matches at $z = 0.07$ ($D=D_h$) to the interior 
fractal-count LTB model  
(for more details of the matching calculation,
see \cite{jmp}).  In all cases the 
significant fact is that the background density $\Omega_{0}$ is 
extremely low.  
Qualitatively, what happens is that the under-density implied
by $N\sim D^\nu$ with $\nu<3$ in the LTB region,
is worsened by the fact that in the corresponding FLRW case,
$N$ has a steep non-power-law gradient beyond $z\approx 0.4$
\cite{rib}.

It is of interest that the above numerical integrations
confirmed that the no-shell-crossing conditions 
\cite{hellaby} were satisfied. The effect of the big-bang function 
$\beta$ on the value of $\Omega_{0}$ is negligible at these 
redshifts. 

 \section{Conclusion}

The nonlinear Hubble law and
low density problems of fractal-count 
universes, identified in this paper, 
apply to all the regular spherically symmetric dust spacetimes 
\cite{jmp}.  These are the spacetimes that can be constructed by 
piecing together regions with LTB metrics 
(including the homogeneous FLRW case), in which the matching 
satisfies the Darmois conditions and there are no surface layers or 
shell crossings. 

The nonlinear Hubble law at very low redshift rules out the
parabolic models.
However,  ways to avoid the low density problem 
of non-parabolic models are still conceivable. There 
could be a surface layer or a cosmological constant, neither of which 
has been included in this analysis, and their effects are not 
fully known.  The most compelling explanation, however, is the 
possibility of a significant bias in the data -- the number counts of 
luminous matter may not trace the actual distribution of density.  
It could be the result of selection effects, evolution or the 
presence of dark matter.  This provides an interesting slant on our 
result because it gives a different emphasis to the search for 
evidence.

 \end{document}